\documentclass[floatfix,aps,twocolumn,amsmath,amssymb,nofootinbib,superscriptaddress,floatfix]{revtex4-1}

\pdfoutput=1

\usepackage{graphicx,color}
\usepackage{float}
\usepackage{amsmath}
\usepackage{amsfonts}
\usepackage[outdir=./]{epstopdf}

\def\newf{\zeta} 

\begin{document}


\title{Correlated rigidity percolation and colloidal gels}


\author{Shang Zhang}
\affiliation{Department of Physics, University of Michigan, Ann Arbor, MI 48109, USA}

\author{Leyou Zhang}
\affiliation{Department of Physics, University of Michigan, Ann Arbor, MI 48109, USA}

\author{Mehdi Bouzid}
\affiliation{Department of Physics, Institute for Soft Matter Synthesis and Metrology, Georgetown University, Washington, D.C. 20057, USA}
\affiliation{LPTMS, CNRS, Univ. Paris-Sud, Université Paris-Saclay, 91405 Orsay, France}

\author{D. Zeb Rocklin}
\affiliation{Department of Physics, University of Michigan, Ann Arbor, MI 48109, USA}
\affiliation{School of Physics, Georgia Institute of Technology, Atlanta, GA 30332, USA}

\author{Emanuela Del Gado}
\affiliation{Department of Physics, Institute for Soft Matter Synthesis and Metrology, Georgetown University, Washington, D.C. 20057, USA}

\author{Xiaoming Mao}
\affiliation{Department of Physics, University of Michigan, Ann Arbor, MI 48109, USA}

\date{\today}

\begin{abstract}
Rigidity percolation (RP) occurs when mechanical stability emerges in disordered networks as constraints or components are added. {Here we discuss RP with structural correlations, an effect ignored in classical theories albeit relevant to many liquid-to-amorphous-solid transitions, such as colloidal gelation, which are due to attractive interactions and aggregation. Using a lattice model, we show that structural correlations shift RP to lower volume fractions. Through molecular dynamics simulations, we show that increasing attraction in colloidal gelation increases structural correlation and thus lowers the RP transition, agreeing with experiments. Hence colloidal gelation can be understood as a RP transition, but occurs at volume fractions far below values predicted by the classical RP, due to attractive interactions which induce structural correlation. } \end{abstract}

\maketitle

\emph{Introduction --}
The emergence of mechanical rigidity in soft amorphous solids is at the core of many material technology developments from 3D printing with soft, biocompatible inks \cite{truby2016printing} to designing food texture \cite{mezzenga2005understanding, keshavarz2017nonlinear}, but it is poorly understood and controlled. The main theoretical framework available is based on the idea that locally rigid structures, due to mechanical constraints such as chemical bonds or steric repulsion, percolate through the material. Hence the problem translates into the onset of rigidity in a disordered network of springs, whose rigidity percolation (RP) transition has been intensively studied, especially in relation with molecular glasses \cite{he1985elastic, jacobs1995generic, sahimi1998non, thorpe2000self, bauchy2011atomic,ellenbroek2015rigidity}. 

With respect to percolation phenomena simply controlled by the geometric connectivity of the involved objects~\cite{stauffer2014introduction}, the onset of rigidity requires a mechanically stable spanning cluster which is able to transmit stresses, a problem intrinsically vectorial and long-range~\cite{jacobs1995generic}, in contrast with the more familiar connectivity percolation.  As a result, RPs occur at {\it higher} transition volume fractions (e.g., $63\%$ for site RP on a two-dimensional triangular lattice~\cite{jacobs1996generic} and $36\%$ for site RP on a three-dimensional face-centered-cubic lattice~\cite{chubynsky2007algorithms})
and display different critical exponents compared to those of geometric percolations.  
{From this perspective, it is surprising that soft amorphous solids such as colloidal gels---formed in suspensions of colloidal particles with prevalently attractive interactions (due to depletion, etc.)---can be mechanically rigid at very low volume fractions, such as a few percent \cite{trappe_nature2001,gisler1999, segre_prl2001}.}  
How can one have a solid at volume fractions far below the predicted RP transition?

In this paper, we address this question by revealing the role of structural correlation on rigidity, especially for the case of attractive particles.  In classical models of RP, bonds or sites are randomly removed from a lattice, with no correlation between one another, until the structure loses its rigidity.  While the classical RP provided well-tested predictions on glass physics, such as the relation between the glass transition temperature and the chemical composition~\cite{he1985elastic}, it ignores any structural correlation between the components.  
It is known that the nature of the rigidity transition can be significantly different, depending on how components of the network are put together~\cite{thorpe2000self,ellenbroek2015rigidity}. For example, when rigidity emerges as frictionless spheres jam due to compression, a spanning rigid cluster suddenly appears that includes nearly all particles in the system and,
 with one more contact, the whole system is stressed~\cite{o2003jamming,liu2010jamming,goodrich2012finite,sussman2016spatial}.  This differs from classical RP where the spanning rigid cluster is fractal at the transition, although both transitions occur near the isostatic point~\cite{liu2010jamming,Mao2010,Ellenbroek2011,Zhang2015a,Lubensky2015}, where the mean coordination number equals two times the spatial dimensions, $\langle z \rangle =2d$. {What makes the emergence of rigidity in jamming so different from the classical RP is that the self-organization of the structure, accommodating  the repulsive interactions among the particles as they are pushed together, dictates the nature of the rigidity transition. It has been recently suggested that the presence of attractive interactions may further change the nature of the rigidity transition at jamming  
 \cite{Koeze2018Sticky-Matter}, but the emergence of rigidity when the self-organization of the structure is due to aggregation and gelation {in a thermodynamic system}~\cite{trappe2001jamming,lois2008jamming,lu2008gelation,Head2007,jorjadze2011attractive} is a much less explored question compared to repulsive particles {being compressed together}, and remains fundamentally not understood.}
 
{For suspensions of attractive colloidal particles
, structural (spatial) correlations are, in most cases, directly accessible in experiments and well rationalized via statistical mechanics approaches: fractal aggregation models, cluster theories and density functional theories can provide good understanding of the structures (and structural correlations) resulting from short-range attractive interactions even at low volume fractions \cite{witten2004structured,hansen1990theory,richard2018coupling}. Nevertheless, gelation has been mainly discussed in terms of the geometric percolation of such structures and of the related particle localization \cite{delgado2004slow,segre_prl2001,broderix2000critical,kroy2004cluster}. Recent work has started to address the rigidity rather than just connectivity of particle aggregates \cite{valadez2013dynamical,zaccone2014linking,Tsurusawa2018Gelation-as-con,Rocklin2018}, but a clear link between colloidal gelation and RP is still lacking.
Here we establish such link by constructing a lattice model} where sites are diluted in a correlated way, mimicking attraction, and we show that the rigidity percolation transition shifts to lower volume fraction as correlation increases, albeit with the same critical exponents as the classical RP (Fig.~\ref{fig:fig1}ab).  Moreover, using molecular dynamics (MD) simulations on a colloidal-gel model where particles aggregate due to short-range attractions, we find that, similar to the lattice model, the presence of spatial correlations due to the particle interactions can lead to RP at progressively lower volume fractions upon increasing the interaction strengths (Fig.~\ref{fig:fig1}cd).  
A simple way to illustrate how structural correlations move the RP down to lower volume fractions is that correlations may organize particles into ``smart'' thin structures that transmit stress.  As shown in Fig.~\ref{fig:fig1}a inset, when particles are arranged on a Warren truss which is rigid, the volume fraction of this one-dimensional structure on a two-dimensional plane vanishes in the thermodynamics limit.  As we show below, attractive correlations naturally prepare particles into these types of structures, giving rise to rigidity at low volume fractions.
\begin{figure}[h]
    \centering
    \includegraphics[width=0.47\textwidth]{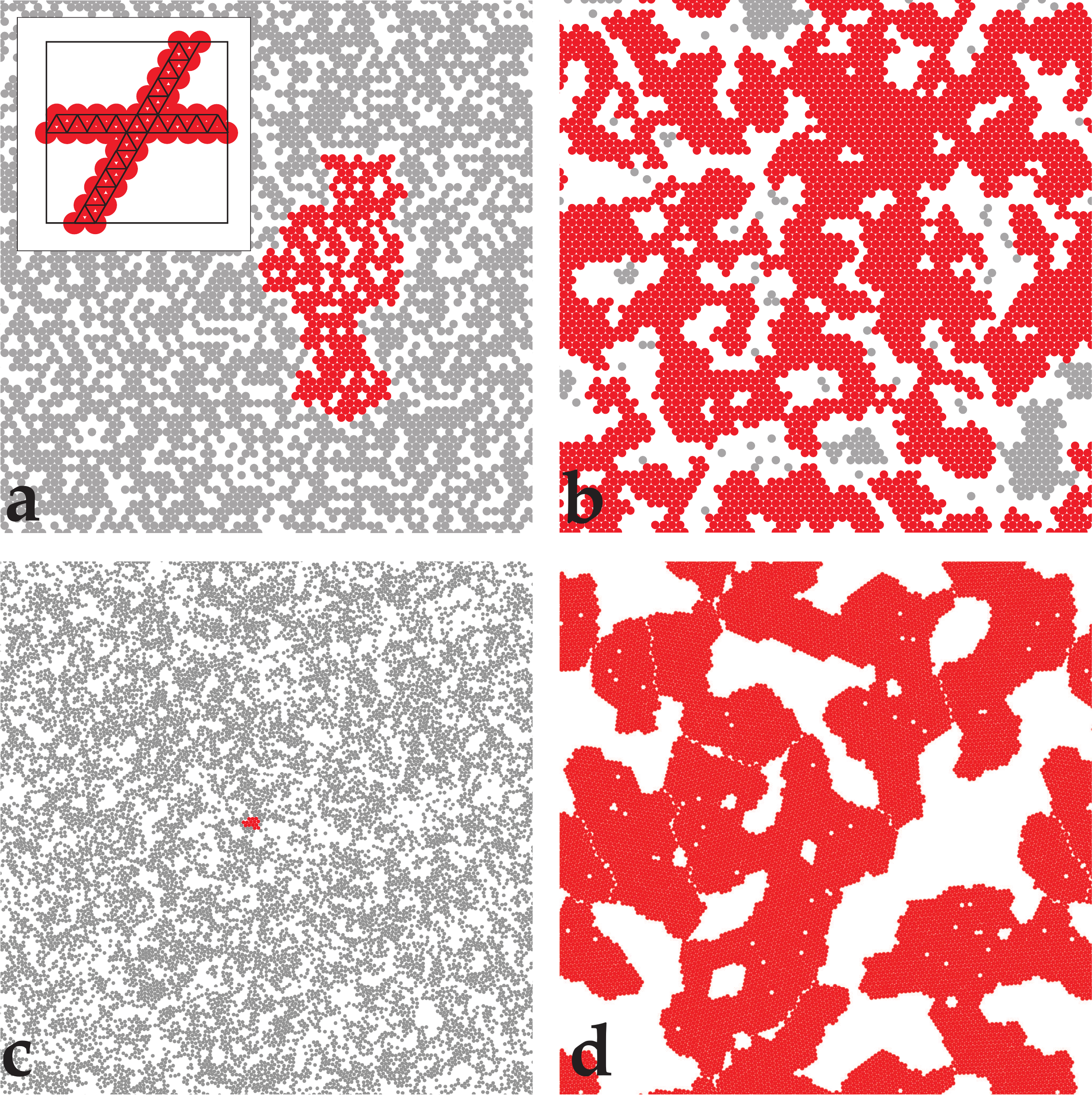}
    \label{fig:fig1}
     \caption{
    Examples of rigid cluster decomposition of the correlated lattice model ($\phi_l  = 0.6$) at different correlation strengths [$c = 0$ in (a) and $c = 0.6$ in (b)], and the attractive gel model ($\phi_{\textrm{g}}=0.6$) at $k_B T/\epsilon=0.4$ in (c) and $0.1$ in (d).  Red particles belong to the largest rigid cluster, and other particles are colored in gray.  
In both the correlated lattice model and the attractive gel model, correlation/attraction induces rigidity at volume fractions below the rigidity transition in the uncorrelated/repulsive limit.  The rigid clusters percolate in (b) and (d) where there is strong correlation/attraction, but not in (a) and (c).  
The inset in (a) shows an extreme example where particles are perfectly correlated (on a Warren truss) and exhibit rigidity at $\phi=0$ in thermodynamic limit.}
\label{fig:fig1}
\end{figure}

\emph{Models and Methods --}
We use two models to investigate the effect of correlation on rigidity.  The first model, which we name the \emph{correlated lattice model}, is a modified version of the site-diluted triangular lattice model for RP~\cite{jacobs1996generic}.  Instead of randomly populating sites in the lattice with a uniform probability, we put particles on a triangular lattice one by one according to the following protocol.  At each step, an empty site is randomly chosen, and a particle is put on this site with probability
\begin{align}\label{EQ:p}
p= (1-c)^{6-N_{nn}}
\end{align}
where $N_{nn}$ is the number of its nearest-neighbor sites which are already occupied ($0\le N_{nn} \le 6$) and $c$ is a dimensionless constant controlling the correlation strength ($0\le c<1$). We start with an empty triangular lattice and repeat this process until a target volume fraction $\phi_{\textrm{l}}$ is reached ({subscript $l$} denotes ``lattice''), which relates to the fraction of occupied sites $f$ through $\phi_{\textrm{l}}\equiv \pi f/(2\sqrt{3})$.  We then obtain a spring network where all nearest neighbor pairs, if both exist, are connected.  
The limit of $c=0$ corresponds to the classical RP with no structural correlation (all sites occupied with the same probability).

The second model, which we name the \emph{attractive gel model}, is an assembly of interacting colloidal particles, studied via molecular dynamics (MD) in 2D.  The particles interact through a pairwise Lennard-Jones-like potential which displays a short range attraction {(of depth $\epsilon$)} and a repulsive core \cite{anderson2002insights,bantawa2018potential}.  We generate configurations at different volume fraction $\phi_g$ (subscript $g$ denotes ``gel''), and different ratios between the thermal energy and the attractive well depth $k_B T/\epsilon$, by solving the many-body Newton's equations of motion in a square simulation box with periodic boundary conditions. For each particle configuration, we obtain the corresponding spring network by assigning bonds between pairs of particles of center-to-center distance $1.03\sigma$ (the inflection point of the potential)  or less.  Further details of our simulation protocol are included in the Supplement Information (SI). We analyze the rigidity of all the spring networks from the two models using the pebble game algorithm~\cite{jacobs1997algorithm,jacobs1995generic}, which decomposes the networks into rigid clusters.  RP occurs when the largest rigid cluster percolates in both directions, leading to macroscopic rigidity~\cite{ellenbroek2015rigidity,henkes2016rigid}.

\begin{figure}[h]
    \centering
    \includegraphics[width=0.45\textwidth]{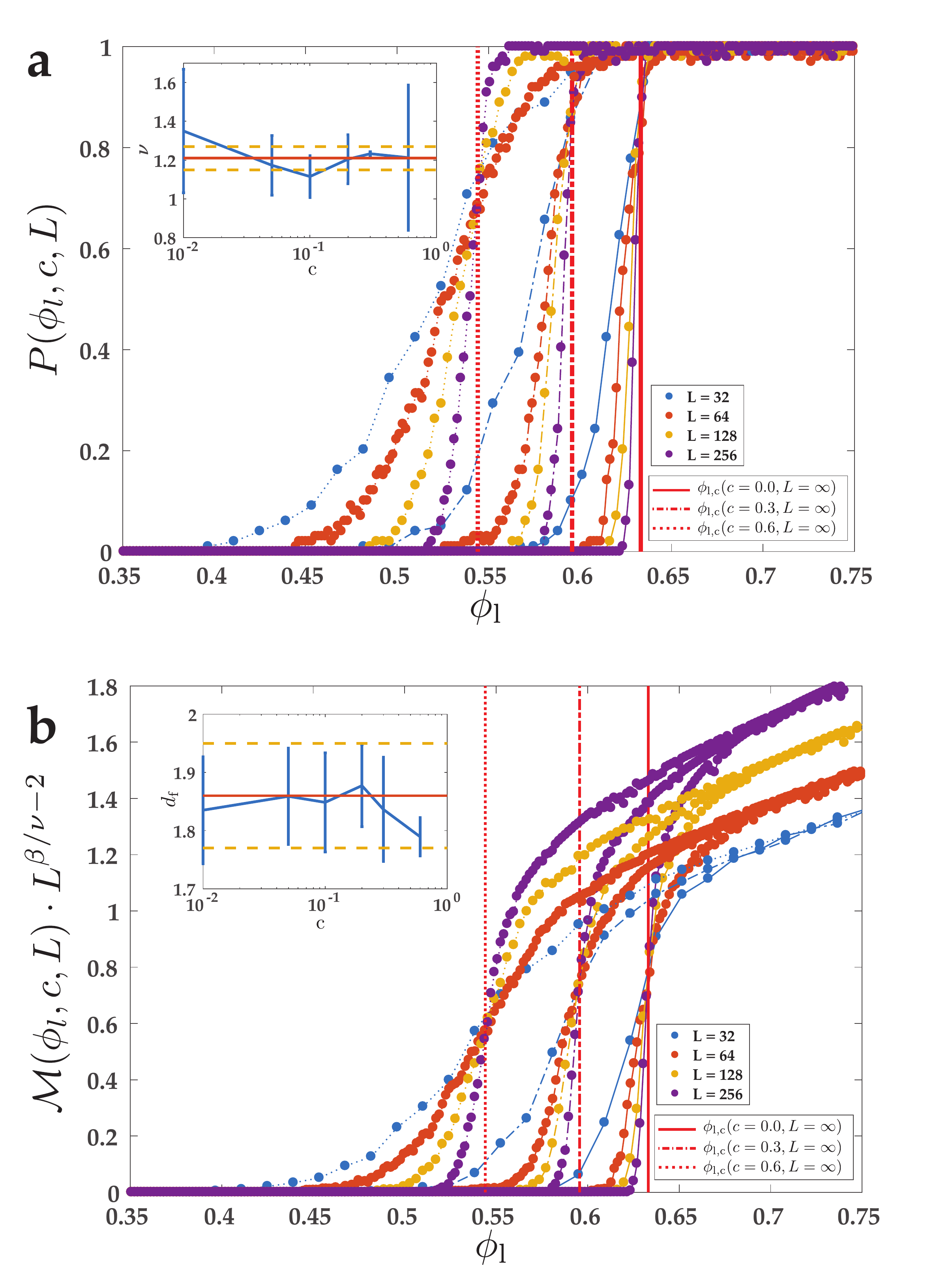}
    \caption{
    (a) $P(\phi_l,c,L)$ at different $L$ and $c$ (symbols and line styles defined in legends).  Inset: $\nu$ for different $c$ (blue with error bars), in comparison with average (red line) and standard error (yellow dashed line) of $\nu$ in the classical RP (from Ref.~\cite{jacobs1996generic}).  
 (b) $\mathcal{M}(\phi_l,c,L)$ at different $L$ and $c$.  Inset: $d_f$ for different $c$ (blue with error bars), in comparison with average (red line) and standard error (yellow dashed line) of $\nu$ in the classical RP (from Ref.~\cite{jacobs1996generic}).
In both (a) and (b), curves for different $L$ cross at the same point (marked by red lines), indicating continuous transitions at every $c$ at different $\phi_{l,c}(c,L=\infty)$.
}\label{fig:fig2}
\end{figure}

\emph{Results --}
In both models, we find that by introducing correlation/attraction, rigidity emerges at lower volume fractions than in uncorrelated cases (Fig.~\ref{fig:fig1}).  
In the correlated lattice model, we measure two quantities, the probability of having a percolating rigid cluster $P (\phi_l, c, L)$, and the average mass of the largest rigid cluster $\mathcal{M}(\phi_l,c, L)$, where $L$ is the linear size  of the lattice. Following the notion of percolation, $\mathcal{M}$ is the order parameter of the transition.  
As shown in Fig.~\ref{fig:fig2}, when the correlation strength $c$ increases, both $P$ and $\mathcal{M}$ curves shift to the left, confirming that RP occurs at a lower $\phi_l$ in the presence of the correlation. 
Moreover, the gradual increase of $\mathcal{M}$ at the transition suggests that the correlated rigidity transition is still continuous, same as the classical RP. The fact that the $P$ and the $\mathcal{M}$ curves at different L intersect at the same scale-free point confirms this.

We further analyze critical scaling relations near the correlated rigidity transition using finite-size scaling (more details in the SI).  We first determine the transition point $\phi_{l,c}(c,L)$ where the spanning rigid cluster first appears, from averaging over disordered samples.  For each $c$, the transition point shifts as a function of $L$ following standard finite-size scaling relations with correlation length exponent $\nu=1.21$ (agreeing with that of the classical RP~\cite{jacobs1995generic}), leading to the infinite volume limit, $\phi_{l,c}(c,L=\infty)$.  We find that the transition point decreases with $c$ following the relation
\begin{align}
	\phi_{l,c}(c=0,L=\infty) - \phi_{l,c}(c,L=\infty) = a\, c^{1/\newf},
\end{align}
at small $c$, where $\newf\simeq 0.76$, the coefficient $a\simeq 0.19$, and the $c\to 0$ limit transition point is $\phi_{l,c}(0,\infty)\simeq 0.63$ agreeing with the classical RP result (note the extra factor of $\pi/(2\sqrt{3})$ converting from site occupancy probability to volume fraction).

We find that our data for $P$ and $\mathcal{M}$ can then be collapsed using the following scaling forms
\begin{align}
\displaystyle P(\phi_l,c, L) &\sim \tilde{P}[(\phi_l-\phi_{l,c}(c,L=\infty))L^{1/\nu}],\\
\displaystyle \mathcal{M}(\phi_l,c,L) &\sim L^{d-\beta/\nu} \tilde{\mathcal{M}}[(\phi_l-\phi_{l,c}(c,L=\infty))L^{1/\nu}],
\end{align}
where $\nu$  and $\beta$ are the critical exponents for the correlation length and the growth of the order parameter {(figures in the SI)}.  These scaling relations share the same form as ones used in classical RP with the same exponents ($\nu=1.21$ and $\beta=0.18$)~\cite{jacobs1995generic}, but with correlation dependent transition points $\phi_{l,c}(c,L=\infty)$ which we determine above.  

Our results suggest that correlations play the role of an irrelevant perturbation at the RP transition.  They shift the transition point $\phi_{l,c}(c,L=\infty)$ while leaving critical exponents the same as in the uncorrelated case. Thus, with correlation, the RP still belongs to the same universality class, as also found in other percolation problems \cite{coniglio1979site,coniglio1979cluster}. One way to interpret this result is that the structural correlations we introduce {in the model are a short range feature}.  Although they shift the transition, the critical {scaling is controlled largely by the physics at large lengthscales and is not sensitive to microscopic modifications}.
We further verify this by measuring the critical exponents directly at different $c$.  In particular, we measure $\nu$ via fluctuations of $\phi_{l,c}(c,L)$ over samples, $\Delta_\phi \equiv \sqrt{\left<\phi_{l,c}(c,L)^2\right>-\left<\phi_{l,c}(c,L)\right>^2}$, as well as the fractal dimension of the giant rigid cluster at the transition $\mathcal{M}_c = \left<\mathcal{M}(\phi_{l,c},c, L)\right>$.  We fit these quantities to their finite-size scaling relations,
\begin{align}
\displaystyle \Delta_\phi &\sim L^{-1/\nu},\\
\displaystyle \mathcal{M}_c &\sim L^{d_f},
\end{align}
where the fractal dimension relates to $\beta$ by $d_f = d - \beta/\nu$ (here $d=2$ is the spatial dimension).  From this analysis we obtain $\nu, \beta$ agreeing with those of the classical RP within error bars for every $c$, as shown in Fig.~\ref{fig:fig2} insets.

\begin{figure}[h]
    \centering
        \includegraphics[width=0.45\textwidth]{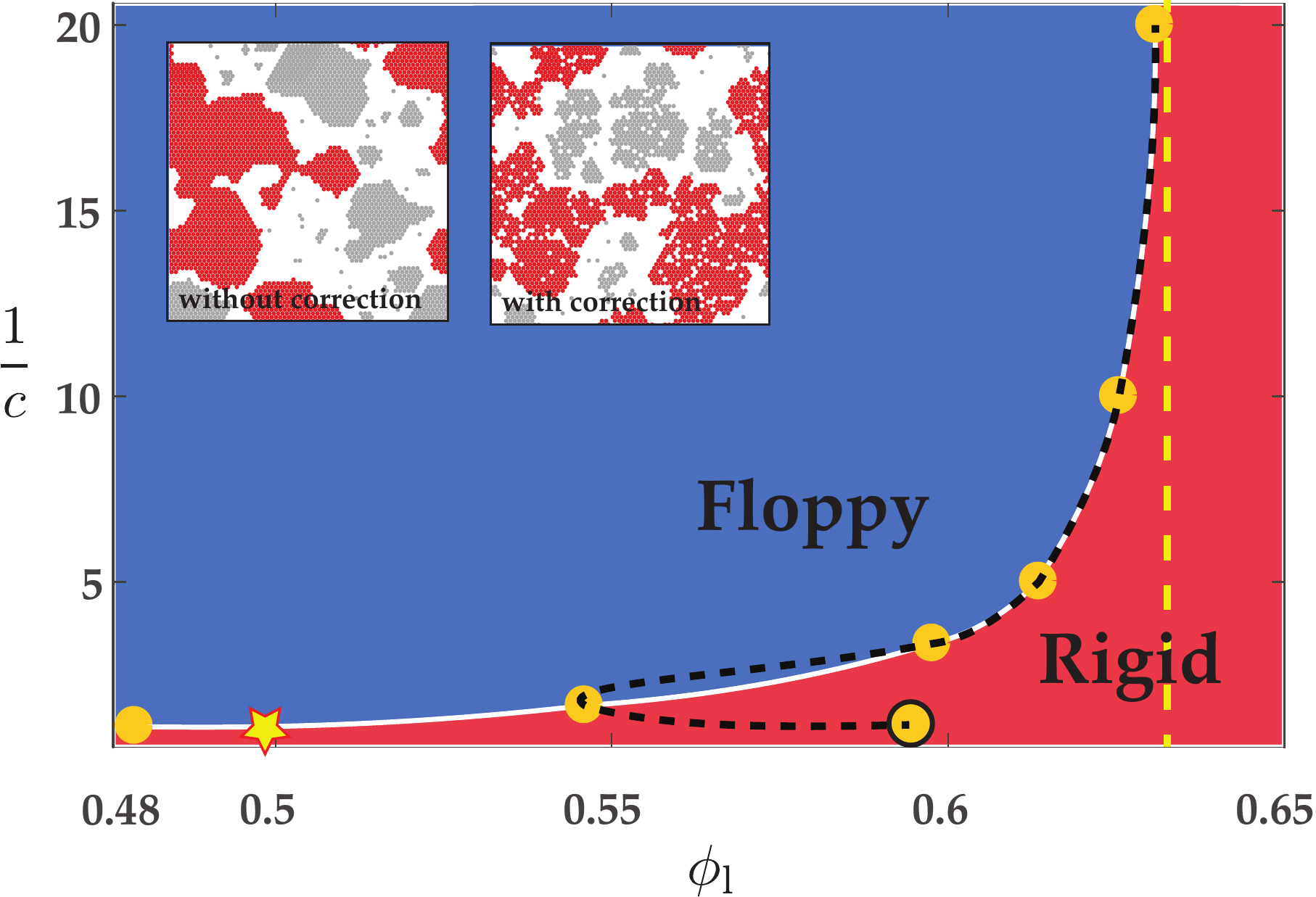}
    \caption{
    Phase diagram of the correlated lattice model (without and with the strong correlation correction).  Calculated phase boundary $\phi_{l,c}(c,L=\infty)$ are shown as yellow dots (before correction: with black circles around the dots, after correction: without circles; They overlap at small $c$).  The black dashed line and the white solid line show the phase boundaries before and after the correction by connecting the dots, respectively.  The $c\to 0$ limit (classical RP) is shown as the yellow dashed line. 
    The insets are configurations taken at $c = 0.9, \phi_l  = 0.5$ (the yellow star) with and without the strong correlation correction, which {avoids the formation of disconnected dense blobs and leads to a percolating rigid cluster.}
}
\label{fig:fig3}
\end{figure}

The resulting phase diagram for the correlated lattice model is shown in Fig.~\ref{fig:fig3}, with the phase boundary determined from our measured $\phi_{l,c}(c,L=\infty)$.  {We plot the phase diagram in the $\phi_l$ vs $1/c$ plane for convenient comparison with the attractive gel model, where we identify the rigid gel states in the $\phi_g$ vs $k_B T/\epsilon$ plane, since correlations decrease as both 
 $1/c$ and $k_B T/\epsilon$ increase.}  In the limit of $1/c \to \infty$ the transition reduces to the classical RP, {while} the boundary shifts to lower $\phi_l$ as $c$ increases {(as discussed above)}.  However, when $c$ is large ($>0.6$) the phase boundary bends back to higher $\phi_l$ (dashed line in Fig.~\ref{fig:fig3}). The reason for this reentrant behavior is that very strong correlations force the particles to aggregate into densely packed blobs that do not percolate. {This high $c$ limit would correspond to an advanced stage of separation of the colloid-dense phase in an attractive colloidal suspension, rather than to the colloidal gelation that takes place through dynamical arrest, preventing the formation of disconnected droplets~\cite{solomon,delgado2010microscopic,zia2014micro}. To better capture such features,} we modify the model by adding a correction for strong correlation to mimic dynamical arrest: a site can not be occupied if 4 or more of its neighboring sites are already occupied [$p=0$ when $N_{nn}\ge 4$ and $p$ still obeys Eq.~\eqref{EQ:p} for $N_{nn}< 4$]. With the modified model, the RP transition volume fraction becomes monotonically decreasing as $c$ increases, in better agreement with experiments and our attractive gel simulation described below.  

\begin{figure}[h]
    \includegraphics[width=0.47\textwidth]{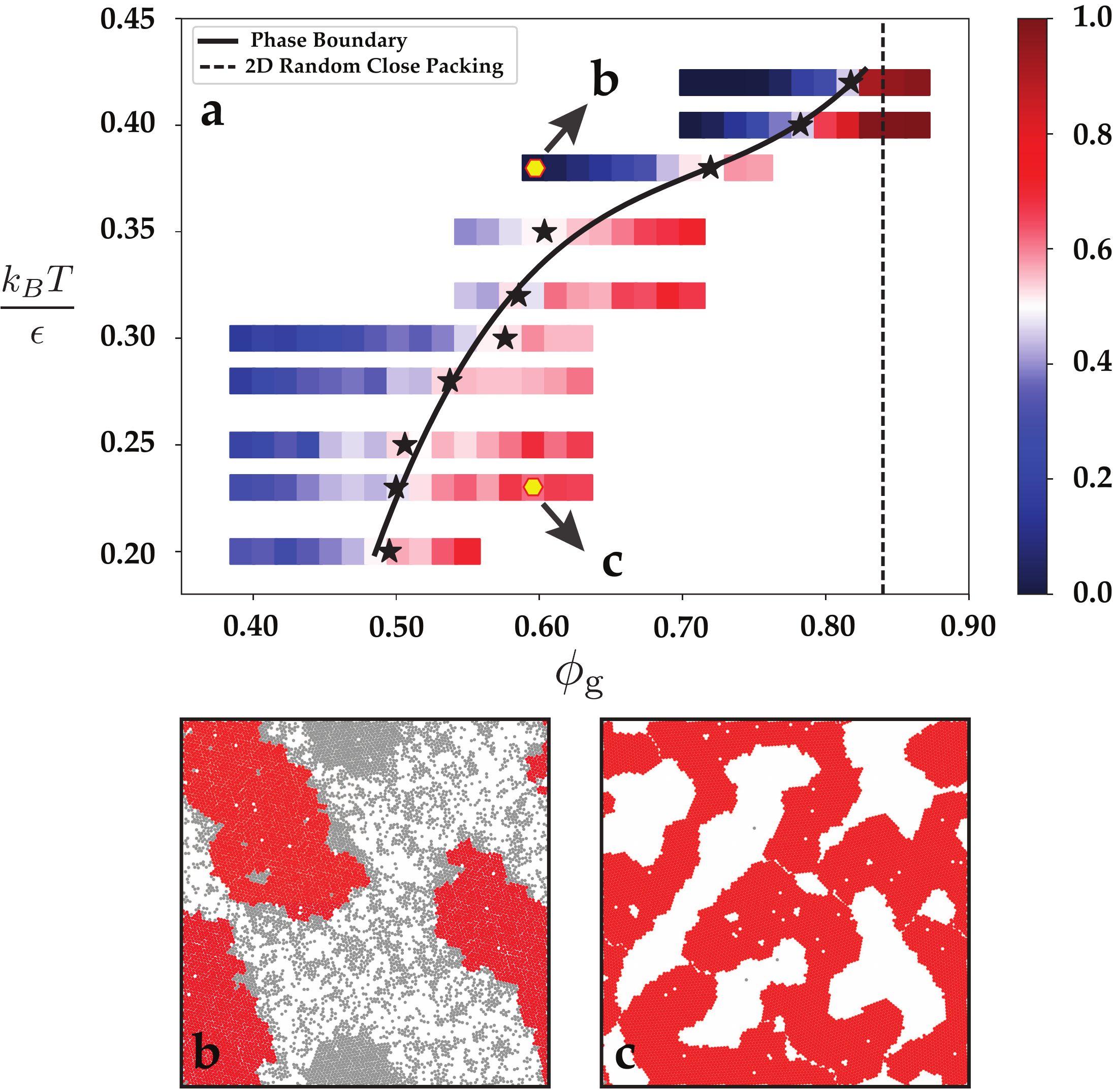}
    \caption{
(a) Phase diagram of the attractive gel model.  Simulated parameters $(\phi_g, k_B T/\epsilon)$ are shown as squares colored according to their measured $P_g(\phi_g, k_B T/\epsilon)$ (color scale shown in legend).  Black stars show fitted phase boundary at each $k_B T/\epsilon$ and the black line is the phase boundary from  fitting  these transition points to third order polynomial. The hard sphere limit of the transition is shown as a black dashed line. (b,c) show two example configurations with their rigid cluster decomposition, chosen at the two marked points on the phase diagram.  The largest rigid cluster percolates in (c) but not in (b), agreeing with the phase boundary.
}
    \label{fig:fig4}
\end{figure}

Results from rigidity analysis of the attractive gel model are shown in Fig.~\ref{fig:fig4}.  We simulate gels of $10^4$ particles in 2$D$ at various $\phi_g$ and $k_B T/\epsilon$, and obtain mean probability for the emergence of a percolating rigid cluster $P_g(\phi_g, k_B T/\epsilon)$. At each $k_B T/\epsilon$ we {identify} the transition point $\phi_{g,c}(k_B T/\epsilon)$ by fitting $P_g(\phi_g, k_B T/\epsilon)$ as a quadratic function of $\phi_g$ and find the point where $P_g=0.5$.  These transition points are then {fitted to a smooth curve to construct} the phase boundary of rigidity in the $\phi_g$ vs. $k_B T/\epsilon$ plane. 
To elucidate the physics of the two phases, we plot two {sample} configurations (with rigid cluster decomposition) at the same volume fraction $\phi_g=0.6$ but for two distinct values $k_B T/\epsilon=0.23$ and $0.38$ (Fig.~\ref{fig:fig4}ab). Large thermal fluctuations are strong enough to frequently break bonds, and the resulting structure is either a homogeneous gas of particles (as shown in Fig.~\ref{fig:fig1}c), or {display phase separation but the large clusters do not show rigidity percolation yet (as shown in Fig.~\ref{fig:fig4}b}). In contrast, decreasing $k_B T/\epsilon$, the attraction is so strong that the particle-rich regions not only exhibit local rigidity, but also percolate through the whole system (as shown in {Fig.~\ref{fig:fig4}c)}. 
{The phase boundary bends down again at very strong attraction, where the system goes out of equilibrium and the rigidity is dominated by the physics of diffusion limited aggregation~\cite{Witten1981}.}
{The similarity between the phase boundaries in the correlated lattice model and the attractive gel model} clearly shows that the rigidity onset in dilute systems is favored by the structural correlations induced by the attractive interactions. Hence colloidal gelation can be understood as a RP transition in which structural correlations help optimize mechanically stable structures \cite{valadez2013dynamical,Tsurusawa2018Gelation-as-con}. Such influence could be further enhanced in the presence of local angular constraints, due to more complex mechanical contacts between colloidal particles \cite{colombo2014self,colombo2014stress}, or under strain \cite{hsiao2012role}.

{To summarize, we have studied} the rigidity transition in a diluted triangular lattice model where particles populate sites with positional correlation, and a colloidal gel model with short range attraction using MD simulation.  The two models show similar structural heterogeneities where particles cluster, forming stress-bearing networks that percolate through the system at low volume fractions.  In particular, we analyze critical scaling exponents in the correlated lattice model, and find that the rigidity transition belongs to the same universality class as the classical RP, {but} the transition point moves to lower volume fractions as correlation increases.
The attractive gel model further demonstrates that such structural correlations and heterogeneities can naturally arise as a result of short range attractive interactions in a thermal system. Deeper understandings of how this structural heterogeneity {develops in the incipient phase separation and how it depends on the preparation protocol used for the gel (for example, the cooling rate in the simulations) \cite{ricateau2018critical}, as well as connecting correlated RP scenario obtained here to the hard sphere limit where no attraction is present and rigidity emerges at the random close packing volume fraction ($84\%$ in 2D) or to the case in which different types of topological  constraints may be present \cite{bouzid2017network}, will be intriguing topics to explore in future studies.}

\emph{Acknowledgements:} We thank M. Solomon for helpful discussions.  
SZ, LZ, and XM thank the support from the National Science Foundation (Grant No. DMR-1609051).  
MB and EDG thank the Impact Program of the Georgetown Environmental Initiative and Georgetown University, Kavli Institute for Theoretical Physics at the University of California Santa Barbara and National Science Foundation (Grant No. NSF PHY17-48958).

\appendix
\section{The ``pebble game'' method: rigid cluster decomposition}

To study the rigidity of the contact networks we obtained, we perform rigidity analysis by decomposing the networks into rigid clusters. The ``pebble game''\cite{jacobs1997algorithm,jacobs1995generic} method is applied.

The ``pebble game'' method is a combinatorial algorithm based on Laman's theorem \cite{laman1970graphs}, which states that a graph with $N$ vertices and $2N-3$ edges is minimally rigid if and only if no subgraph of $n$ vertices has more than $2n-3$ edges. Laman's theorem counts constraints beyond the mean-field theory in two dimensions. The ``pebble game'' method is an efficient way to apply Laman's theorem to networks and is able to perform tests such as (i) calculating the number of floppy modes, (ii) identifying over-constrained regions and (iii) locating rigid clusters.

For a given contact network obtained from the correlated lattice model or the attractive gel model, we assign each particle $d$ pebbles that match its $d$ degrees of freedom, where $d=2$ is the dimension. We then use the ``pebble game'' method to classify each contact as either an independent constraint that absorbs one pebble, or a redundant constraint that absorbs no pebble. Rigid clusters, subsets of the system where contacts absorb all degrees of freedom except the $d(d+1)/2$ rigid body motions, are then identified. 

We can determine whether the largest rigid cluster spans around the system by testing whether the cluster wraps around the periodic boundary of the lattice~\cite{machta1996invaded,newman2001fast}. Fig.~\ref{fig:fig1} shows example configurations of our correlated model as well as attractive gel model, with rigid clusters identified by the ``pebble game'' algorithm marked.


\section{Colloidal gel simulation}

Our model for the colloidal system is a $2D$ assembly of of $N=10^4$ particles (monodisperse in size) interacting through a Lennard-Jones-like potential 
\begin{equation}
U(\mathbf{r}_1,\cdots,\mathbf{r}_N) = \displaystyle \epsilon \sum_{i>j} u(\frac{\mathbf{r}_{i}-\mathbf{r}_{j}}{\sigma})
\end{equation}
where $\epsilon$ is the potential energy scale (setting the unit energy in our simulations) and $\sigma$ is the diameter of particles (setting the unit length). 
$u(\mathbf{r})$ is a potential well obtained by combining, in the spirit of Lennard-Jones potential, an attractive term with a short range repulsive core, and, for computational convenience, is written as
\begin{equation} 
u(\mathbf{r}) = A(a\ r^{-18}-r^{-16}),
\end{equation}
where $\mathbf{r}$ is the interparticle distance rescaled by the particle diameter and $A$ and $a$ are dimensionless constants. In particular we have fixed $A = 6.27$, $a = 0.85$ to obtain a short-ranged attractive well of depth $\epsilon$ and range $\simeq 0.3 \sigma$ \cite{colombo2014self,colombo2014stress}. We adopt periodic boundary conditions and, using the particle diameter $\sigma$, we define an approximate volume (surface) fraction $\phi_g = \pi (\sigma/2)^{2} N/ L^{2}$, where $L$ is the side length of the square simulation box (in units of $\sigma$). We then set the box length according to the target volume fraction $\phi_g$. The gel configurations are obtained using Molecular Dynamics (MD) and a Nos\'e-Hoover thermostat to control the temperature \cite{frenkel2002understanding}, to mimic different interaction strengths $k_{B}T/\epsilon$ as usually done when simulating interacting colloidal particles \cite{noro2000extended,anderson2002insights}. For the gel preparation we solve Newton's equations of motion for computational efficiency, having checked that the gel configurations obtained through the procedure described below do not meaningfully vary with varying the microscopic dynamics (i.e. Newton's {\it vs.} Langevin overdamped dynamics). For the MD simulations we use a time step $\delta t =0.005 \tau_{0}$, where $\tau_{0}=\sqrt{m \sigma^{2}/\epsilon}$ is the usual MD unit time ($m$ is the particle mass). All simulations reported here have been performed with LAMMPS \cite{plimpton1995fast}, suitably modified by us to include the interactions above.

The particles are initially equilibrated at a high temperature ($T \simeq 1$ in units of $\epsilon/k_B$) and then slowly quenched to different target temperatures, corresponding to different $k_{B}T/\epsilon$ values, for $2\cdot 10^{6}$ MD steps. For the lowest target temperatures, $T\le0.32\epsilon/k_B$, we make sure the system has reached a local minimum of the potential energy by solving the damped equations of motion 
\begin{align}
\displaystyle m \frac{d^2\mathbf{r}_i}{dt^2} &= -\xi \frac{d\mathbf{r}_i}{dt} - \Delta_{\mathbf{r}_i} U,
\end{align}
where $\xi$ is the damping coefficient and has units of $m/\tau_{0}$, for $2\cdot 10^{5}$ MD steps, within which the kinetic energy of the system drops to $\simeq 10^{-10} \epsilon$. All data discussed here have been averaged over 200 independently generated samples.

\section{Finite-size scaling for lattice model}


%


\subsection{Identifying the position of phase transition boundary}

 We randomly generate 100 realizations of the correlated lattice model for each correlation strength $c$ and system size $L$ and identify the critical volume fractions for each configuration. The average critical volume fractions of the ensembles are measured by fitting a Gaussian distribution $\mathcal{N}(\mu,\sigma^2)$ to the probability distribution of the critical volume fractions, i.e. $\phi_{\text{l,c}}(c,L) = \mu$. 
 
\begin{figure}[H]
    \centering
    \includegraphics[width=0.4\textwidth]{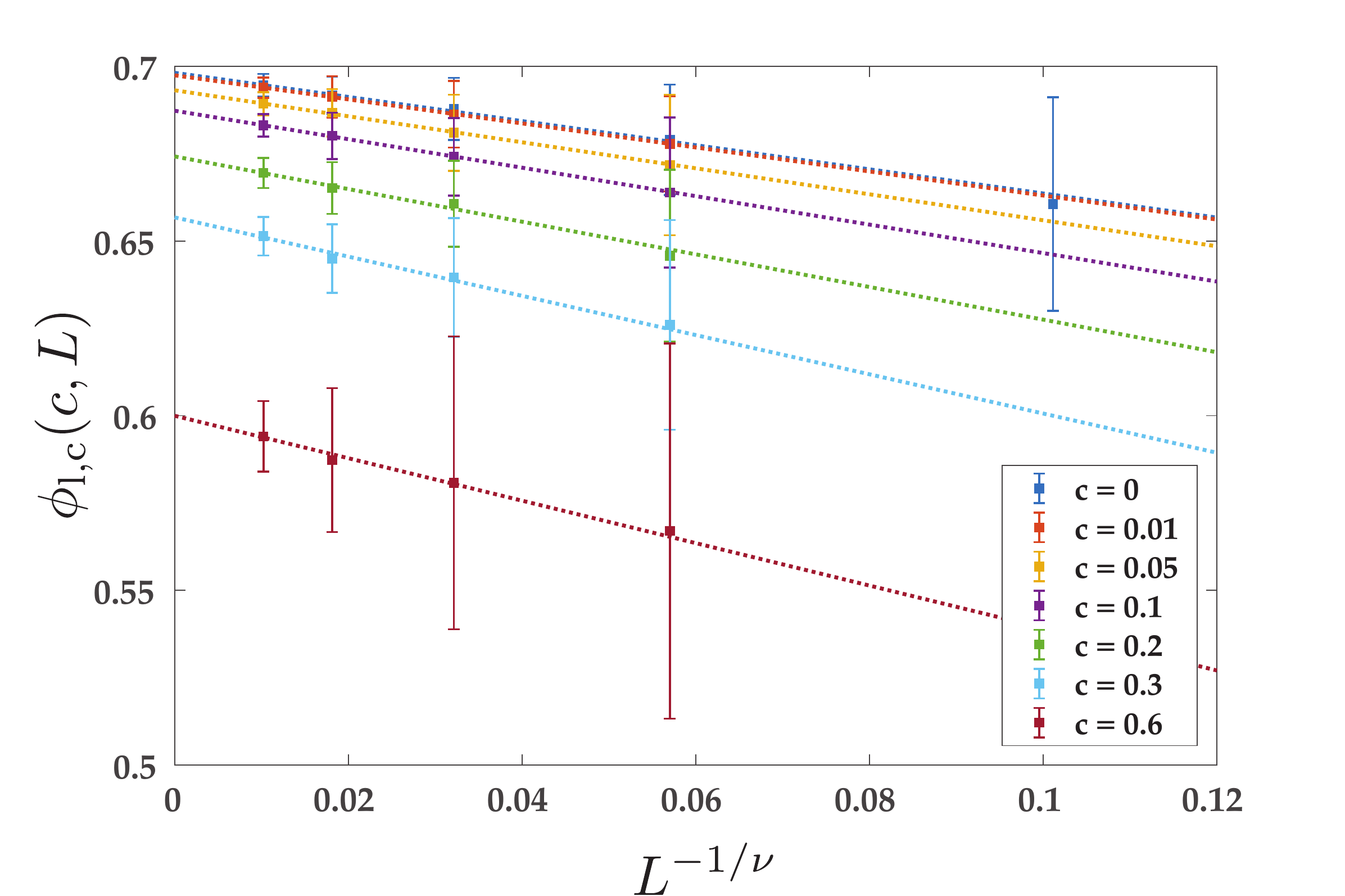}
    \caption{Linear extrapolation of the finite-size critical volume fractions as a function of $L^{-1/\nu}$ ($\nu = 1.21$ \cite{jacobs1995generic}). The dashed lines are linear fits, and the y-intersects of these dashed lines represent the infinite-size limit of the critical volume fractions $\phi_{\text{l,c}}(c,L = \infty)$.}
    \label{fig:finite}
\end{figure}

We then linearly extrapolate these finite critical volume fractions $\phi_{\text{l,c}}(c,L)$ for each $c$ as a function of $L^{-1/\nu}$ to obtain the infinite-size limit $\phi_{\text{l,c}}(c,L=\infty)$ from the y-intersects of the linear fits, as shown in Fig.~\ref{fig:finite}. Here, we adopt $\nu=1.21$ from the uncorrelated RP. We also  directly measure the critical exponents in the correlated RP (section III.C), which are shown to be the same within error bars as those in the uncorrelated RP .


\begin{figure}[H]
    \centering
    \includegraphics[width=0.4\textwidth]{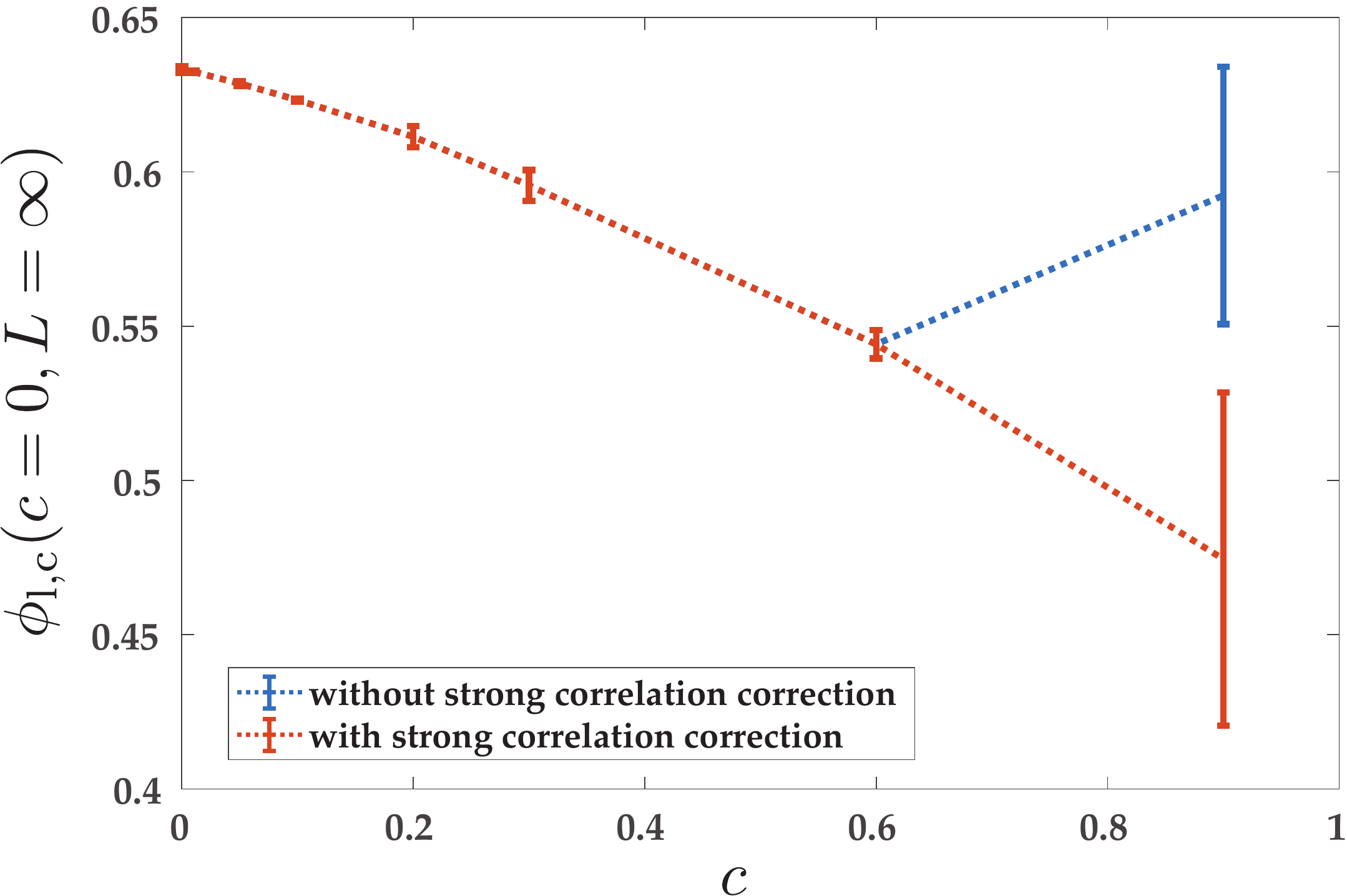}
    \caption{Transition point $\phi_{\text{l,c}}(c=0,L=\infty)$ as a function of $c$ for the correlated lattice model, without and with the strong correlation correction.}
    \label{fig:phase}
\end{figure}

Using the extrapolated infinite-size critical volume fractions, we obtain the phase boundary for the correlated RP, as shown in Fig~\ref{fig:phase}a. As a comparison, we also show the phase boundary of the modified model that is introduced in the main text (Fig~\ref{fig:phase}b). In this phase diagram, all the critical volume fractions are also in the infinite-size limit. The only visible difference of the two phase boundaries is at $c=0.9$.

Scaling behavior of $\phi_{\text{l,c}}(c=0,L=\infty)$ as a function of $c$ at small $c$ is shown as
\begin{align}
\displaystyle |\phi_{\text{l,c}}(c=0,L=\infty) - \phi_{\text{l,c}}(c,L=\infty)| = a\, c^{1/\newf}.
\label{eq:ph_boundary_scaling}
\end{align}
The fitted parameters are $\newf \approx 0.76$ and $a\simeq 0.19$. This fitting provides a quantitative measurement for the magnitude of the critical volume fraction shift as correlation strength $c$ increases. 

\begin{figure}[H]
    \centering
    \includegraphics[width=0.4\textwidth]{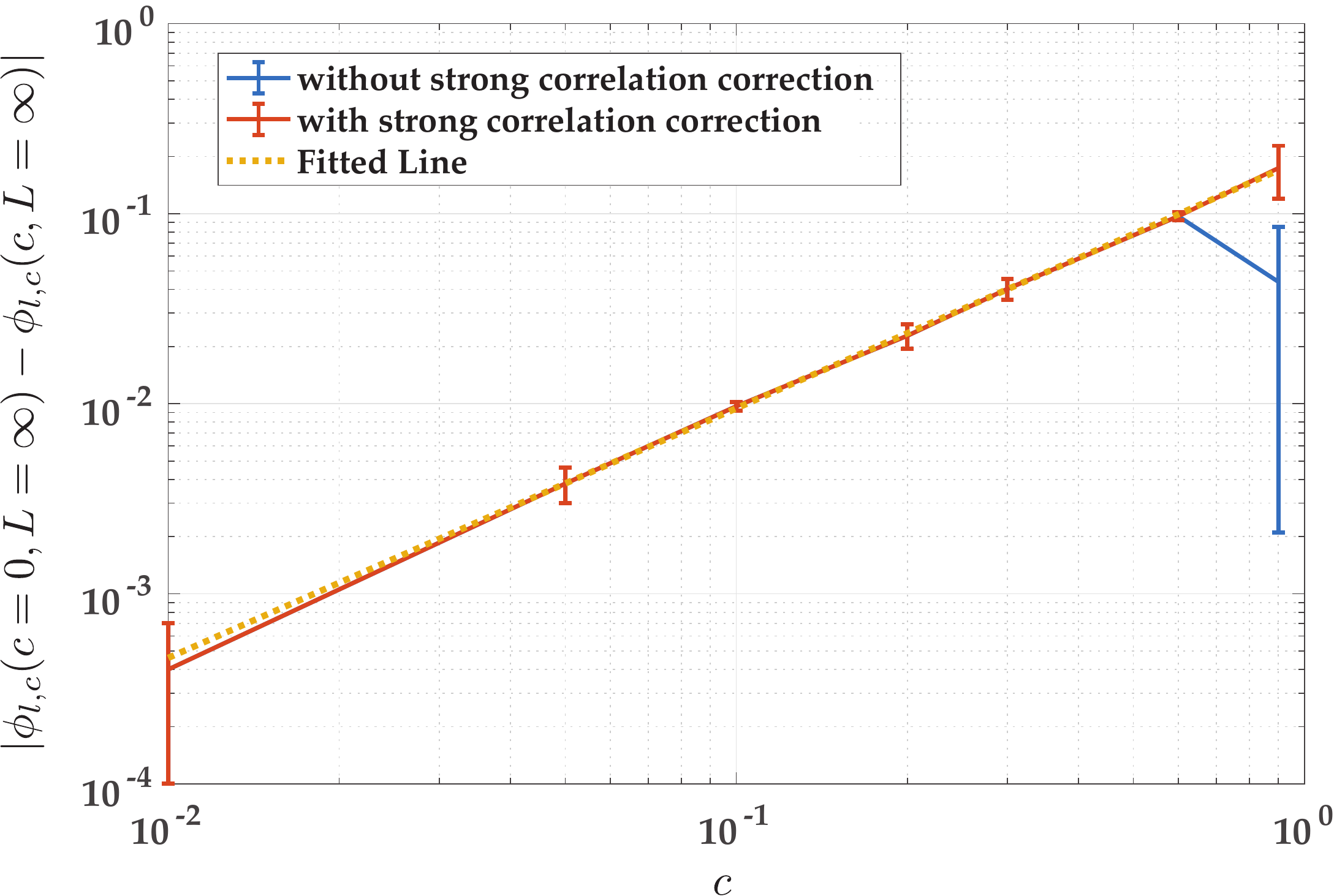}
    \caption{The fitting of the phase boundary as a function of the correlation strength $c$, as described in Eq.~(\ref{eq:ph_boundary_scaling}), without and with the strong correlation correction.}
    \label{fig:PBScaling}
\end{figure}
\subsection{Scaling forms and data collapse}

\begin{figure}[H]
    \centering
    \includegraphics[width=0.4\textwidth]{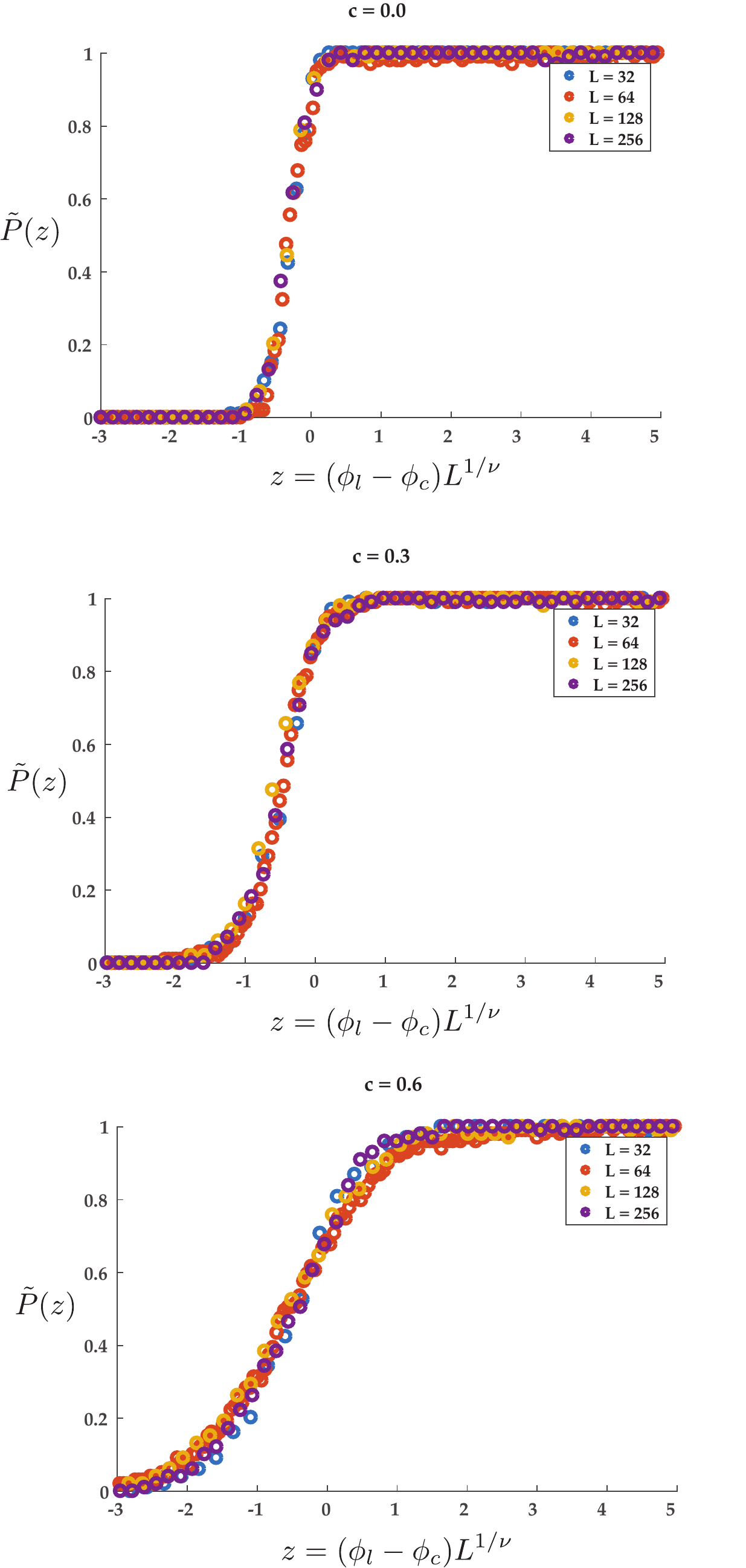}
    \caption{Collapsing of data for $P$, the probability of having a spanning rigid cluster, for the correlated RP, showing the master curve $\tilde{\mathcal{P}}(z)$, where $z=[\phi_l-\phi_{\text{l,c}}(c,L=\infty)]L^{1/\nu}$ and $\nu=1.21$. The exponent is from the uncorrelated RP\cite{jacobs1995generic}).}
    \label{fig:PrScale}
\end{figure}

\begin{figure}[H]
    \centering
    \includegraphics[width=0.4\textwidth]{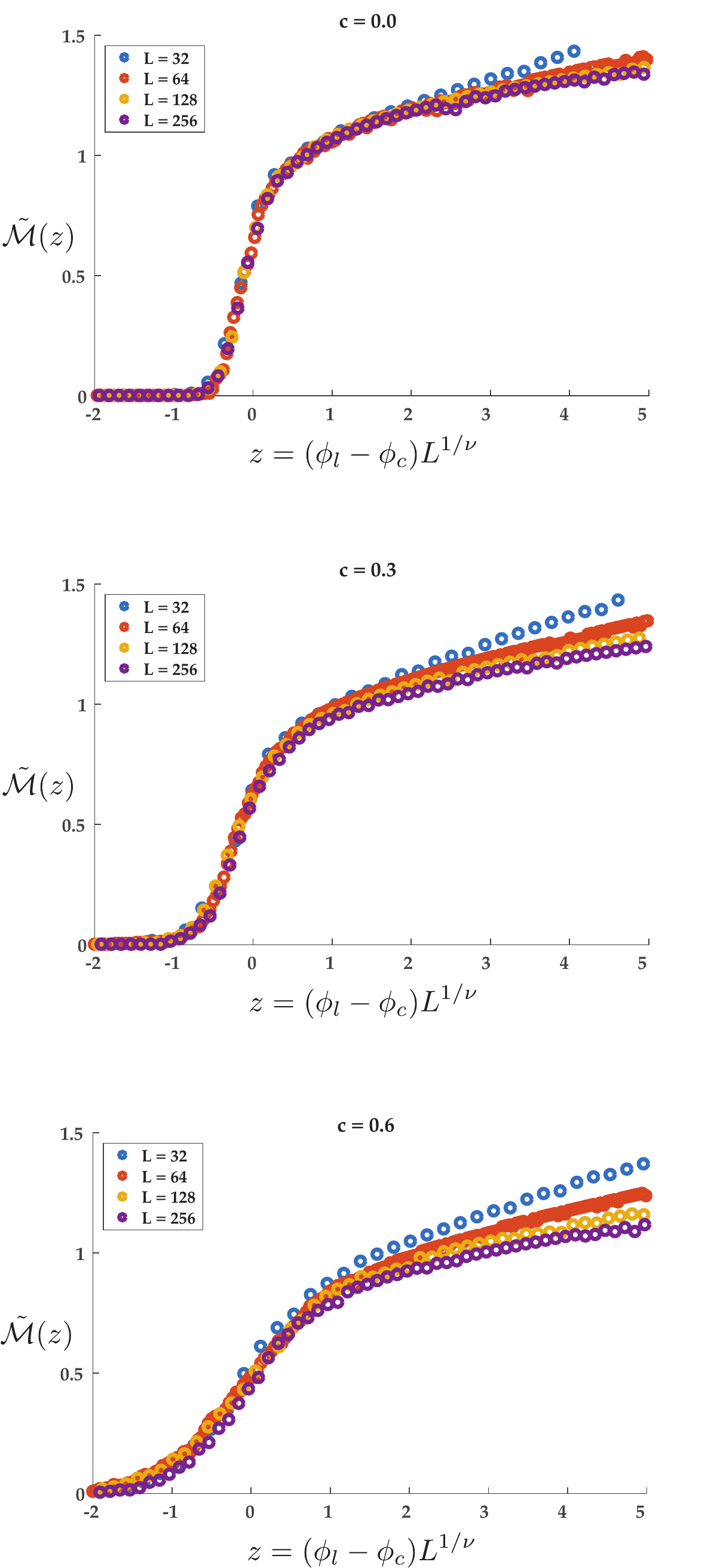}
    \caption{Collapsing of data for $\mathcal{M}$, the mass of rigid cluster in the correlated RP, showing the master curve $\tilde{\mathcal{M}}(z)$, where $z = (\phi_l-\phi_{\text{l,c}}(c,L=\infty))L^{1/\nu}$ and $\beta=0.18$, $\nu=1.21$. The exponents are from the uncorrelated RP~\cite{jacobs1995generic}.}
    \label{fig:massScale}
\end{figure}

The scaling functions of the rigidity percolation are described as:
\begin{align}
\displaystyle P(\phi_l,c, L) &\sim \tilde{P}[(\phi_l-\phi_{\text{l,c}}(c,L=\infty))L^{1/\nu}],
\end{align}
and 
\begin{align}
\displaystyle \mathcal{M}(\phi_l,c,L) &\sim L^{d-\beta/\nu} \tilde{\mathcal{M}}[(\phi_l-\phi_{\text{l,c}}(c,L=\infty))L^{1/\nu}].
\end{align}
Here, we show that using the extrapolated infinite-size critical volume fractions and the critical exponents $\nu$ and $\beta$ adopted from the uncorrelated RP, we are able to collapse the data of $P$ and $\mathcal{M}$. (Fig.~\ref{fig:PrScale} and Fig.~\ref{fig:massScale}).

\subsection{Direct measurement of critical exponents}

To further verify that the correlated RP is in the same universality class as the uncorrelated RP, we also directly measure the critical exponents in the correlated RP from the following scaling relations 
\begin{align}
\displaystyle \Delta_\phi &\sim L^{-1/\nu},
\label{scalingeq1}
\end{align}
and
\begin{align}
\displaystyle \mathcal{M}_c &\sim L^{d_f},
\label{scalingeq2}
\end{align}
where $d_f = 2-\beta/\nu$.

The critical fluctuation $\Delta_\phi$ of critical volume fraction is measured as the standard deviation $\sigma$ of the Gaussian distribution $\mathcal{N}(\mu,\sigma^2)$ fitted from the probability distribution of $\displaystyle P(\phi_l,c, L)$ for each $c$ and $L$.  The critical exponent $\nu$ is then fitted using Eq.~(\ref{scalingeq1}). The fractal dimension for the spanning rigid cluster $d_f$ is fitted from the finite-size scaling of $\mathcal{M}_{\text{c}}(\phi_l,c,L)$ for each $c$ and $L$ using Eq.~(\ref{scalingeq2}).  The exponent $\beta$ are then obtained using $d_f = 2-\beta/\nu$.

\section{The rigidity diagram for gelation of colloidal particles}

\subsection{Defining contact networks}

To construct the rigidity phase diagram of the colloidal particles, we need to define the spring network in the configurations generated through the MD simulations and then perform the ``pebble game'' algorithm to identify rigid clusters. We assume that a spring can be places between two particles that are separated within a distance close to the minimum of the potential well. {In our case, we choose $ \simeq 1.03 \sigma$, which is the inflection point of the interaction potential, defined as the distance where the second derivative of the potential is zero, i.e. 
\begin{align}
\displaystyle U^{''}(r) = 0.
\end{align}
At the inflection point, the attractive force starts to decrease with distance.} At such distance two particles can be considered as bonded and the interactions can be approximated as a spring and hence $1.03 \sigma$ is considered as the bond length in our analysis.

At low temperatures, bonded particles are indeed separated within this range in the vast majority of cases and hence using different lengths to identify the contact, or the bonded state (e.g. the inflection point or the total range of the attractive well) does not affect the results obtained. At high temperatures, instead, particles distances can vary significantly even when they are persistently within the well range, due to their kinetic energy, and they don't necessarily sit in the potential minimum.  Nevertheless, the overall analysis of the rigidity boundary does not significantly change when we consider different bond lengths within the well range, due to the fact that we average the results of the rigidity analysis over several {different MD initial configurations}: either particles are instantaneously interacting but the bond between them is not persistent (and hence not relevant for the rigidity analysis) or the particles are actually interacting over a finite time and the potential minimum is indeed the most probable interparticle distance. For one specific target temperature, we generate 200 different initial configurations, and for each initial configuration we can extract the spring network using the bond length range 1.03$\sigma$. Then we can average over these results of the rigidity analysis to get the probability of having a spanning rigid cluster.

\begin{figure}[H]
\centering
\includegraphics[width=0.4\textwidth]{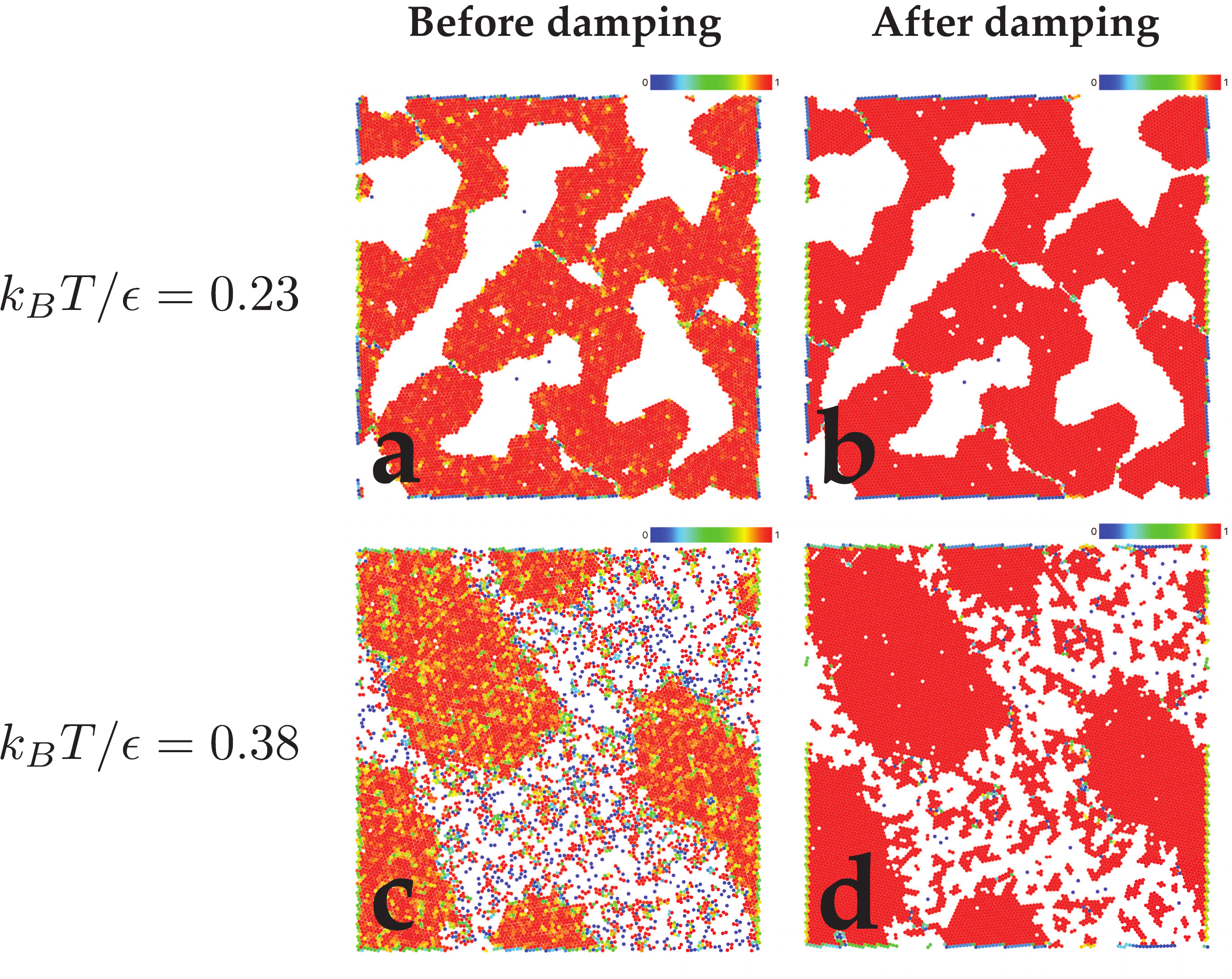}
\caption{Bond orientational order parameter (BOP) in sample configurations. (a,b) $\displaystyle {k_B T}/{\epsilon} = 0.23, \phi = 0.6$. (c,d) $\displaystyle {k_B T}/{\epsilon} = 0.38, \phi = 0.6$. (a,c) BOPs without the damping process. (b,d) BOPs with the damping process.
}
\label{fig:fig_phi6}
\end{figure}

We also test the persistence of the clusters of bonded particles using the damping procedure described above. The local structures have been characterized in terms of the local bond orientational order parameter (BOP) characterizing local crystalline order in 2D~\cite{steinhardt1983bond}. The BOP of a particle $k$ is defined as:
\begin{align}
\psi_6(k) &= \displaystyle \frac{1}{N} \sum_{l}^{N} \exp(i\cdot 6 \theta_{l}),
\end{align}
where $N$ is the number nearest neighbors for particle $k$ and $\theta_l$ is the direction of neighboring particle $l$. Dense clusters of particles tend to be locally crystalline since the particles have the same size. (Fig.~\ref{fig:fig_phi6}ab) shows, as expected, that for low temperatures ($\displaystyle {k_B T}/{\epsilon} \leq 0.32$ in our simulation), the damping tends to preserve the local structures of rigid clusters, and the obtained contact networks reflect the rigidity of the low temperature systems. For high temperatures ($\displaystyle {k_B T}/{\epsilon} > 0.32$ in our simulation), instead, the damping modifies the aggregates local structure, since those aggregates tend not to  persist over time (Fig.~\ref{fig:fig_phi6}cd). 

\subsection{Rigidity boundary for the colloidal gel model}

To find the transition point for RP of colloidal particles, we fit the mean probability for the emergence of a percolating rigid cluster $P_g(\phi_g, k_B T/\epsilon)$ to quadratic functions and interpolate for $P_g(\phi_{\text{g,c}}, k_B T/\epsilon) = 0.5$. These transition points at each $k_B T/\epsilon$ are shown in Fig.\ref{fig:fig4} and used to fit the phase boundary.

\begin{figure}[H]
\centering
\includegraphics[width=0.4\textwidth]{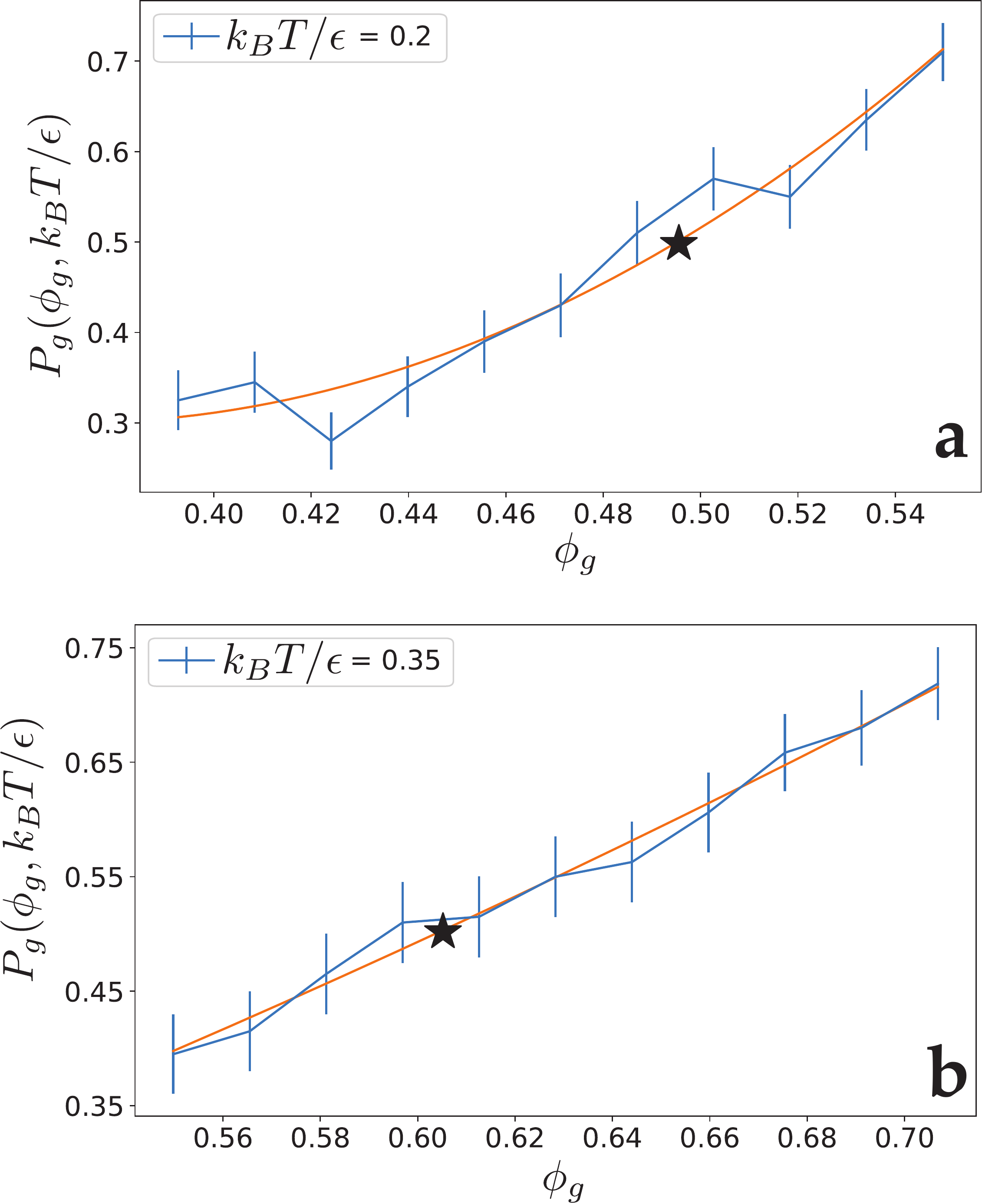}
\caption{Fitting of $P_g(\phi_g, k_B T/\epsilon)$ to second order polynomial in two example cases, 9a)$k_BT/\epsilon = 0.2$ and (b)$k_BT/\epsilon = 0.35$, where the black stars indicate the extrapolated transition point defined as $P_g(\phi_g,k_BT/\epsilon)=0.5$.}
\label{fig:fig_MDfitting}
\end{figure}

In Fig.~\ref{fig:fig_MDfitting} we show two examples $P_g(\phi_g, k_B T/\epsilon)$ fitted to second order polynomial functions. The interpolated transition points $\phi_{\text{g,c}}$ are also marked.

%

\end{document}